\documentclass{article}
\usepackage[preprint]{corl_2023} 
\usepackage{graphicx}
\usepackage{booktabs} 
\usepackage{longtable} 
\usepackage{array} 
\title{HetaRAG: Hybrid Deep Retrieval-Augmented Generation across Heterogeneous Data Stores}
\author{
  Guohang Yan, Yue Zhang, Pinlong Cai$^{*}$, Ding Wang, Song Mao, Hongwei Zhang \\ 
  \textbf{Yaoze Zhang, Hairong Zhang, Xinyu Cai, Botian Shi\thanks{Corresponding authors}}
 \\
  \\Shanghai Artificial Intelligence Laboratory, Shanghai, China
}
\begin{document}
\maketitle


\begin{abstract}
Retrieval-augmented generation (RAG) has become a dominant paradigm for mitigating knowledge hallucination and staleness in large language models (LLMs) while preserving data security. By retrieving relevant evidence from private, domain-specific corpora and injecting it into carefully engineered prompts, RAG delivers trustworthy responses without the prohibitive cost of fine-tuning. Nevertheless, existing systems remain brittle when confronted with (i) multimodal evidence that spans text, images, and structured graphics; (ii) complex queries requiring compositional reasoning over long, noisy contexts; and (iii) continuously evolving knowledge scattered across heterogeneous corpora. 
Traditional retrieval-augmented generation (RAG) systems are text-only and often rely on a single storage backend—typically a vector database. In practice, this monolithic design suffers from unavoidable trade-offs: vector search captures semantic similarity yet loses global context; knowledge graphs excel at relational precision but struggle with recall; full-text indexes are fast and exact yet semantically blind; and relational engines such as MySQL provide strong transactional guarantees but no semantic understanding. We argue that these heterogeneous retrieval paradigms are complementary, and propose a principled fusion scheme to orchestrate them synergistically, mitigating the weaknesses of any single modality.
In this work we introduce HetaRAG, a hybrid, deep-retrieval augmented generation framework that orchestrates cross-modal evidence from heterogeneous data stores. HetaRAG handles multimodal documents—text, diagrams, tables, mathematical notation, and more—and performs sophisticated, compositional reasoning over long-form, noisy contexts. 
We plan to design a system that unifies vector indices, knowledge graphs, full-text engines, and structured databases into a single retrieval plane, dynamically routing and fusing evidence to maximize recall, precision, and contextual fidelity.
To achieve this design goal, we carried out preliminary explorations and constructed an initial RAG pipeline; this technical report provides a brief overview.
The partial code is available at 
\url{https://github.com/KnowledgeXLab/HetaRAG}.
\end{abstract}



\section{Introduction}

With the rapid development of information technology and the explosion of data volume, efficiently acquiring and utilizing knowledge has become crucial for the advancement of intelligent systems. A knowledge engine, as a decision-support system that integrates data, models, and inference rules, plays an essential role in transforming unstructured and non-centralized data into structured knowledge, enabling effective information retrieval, reasoning, and decision-making in specific domains.


In recent years, large language models have demonstrated remarkable capabilities in natural language understanding and generation. However, when applied to specialized domains or high-precision tasks, these models often lack sufficient factual grounding, resulting in unreliable or inaccurate outputs. Therefore, integrating external knowledge sources with large language models has become essential to enhance their practical applicability. By leveraging a knowledge engine capable of providing accurate and multimodal information, large language models can better perform information extraction, retrieval matching, and content generation—ultimately improving the efficiency and quality of knowledge processing.


Despite the availability of several knowledge engine products, these systems still exhibit notable limitations. For instance, they are primarily designed for individual users uploading small volumes of documents, with limited scalability in document processing. Moreover, most rely on vector similarity matching and keyword-based retrieval, which often result in low precision. Their multimodal processing capabilities remain rudimentary, failing to meet the growing demand for cross-modal integration. Modern knowledge bases must go beyond simple storage and retrieval functions. They should support advanced features such as knowledge reasoning, dynamic updating, and data privacy protection to cater to diverse application scenarios across industries. Additionally, the construction and optimization of knowledge bases are ongoing processes that require continuous user feedback and iterative improvements.
    
In this technical report, we present HetaRAG, a hybrid, deep-retrieval RAG framework that unifies multiple heterogeneous data stores—vector indices, knowledge graphs, full-text search engines, and relational databases. The pipeline proceeds in two phases. First, we perform multimodal document ingestion that extracts text, images, tables, and mathematical formulae from raw files. Each modality is then transformed into the most appropriate storage format: dense vectors for semantic proximity, symbolic triples for relational precision, inverted indices for lexical recall, and normalized tables for transactional integrity. Second, at generation time, the system orchestrates evidence from these disparate stores to answer user questions or synthesize comprehensive, multi-modal research reports. The open-source release of HetaRAG currently offers the following core features: 
    
    \begin{itemize}
        \item \textbf{Multimodal document parsing}: decomposing the input document into semantically distinct segments—text, images, tables, mathematical formulas, etc.
        \item \textbf{Heterogeneous storage index construction}: simultaneous indexing into vector databases, knowledge graphs, full-text engines, and relational databases.
        \item \textbf{Deep research report generation}: transforms unstructured documents into verified, query-aligned Markdown reports by fusing text, tables, and visuals via LLM synthesis and vector retrieval.
        \item \textbf{Complex question answering via iterative, reasoning}: iteratively rewrites queries and retrieves evidence until a coherent, cross-source answer emerges.
    \end{itemize}



\section{Related Work}

\subsection{Bottlenecks and Drivers for RAG System Improvement}
With the development of large language models (LLMs), retrieval-augmented generation (RAG) has become a crucial approach to addressing knowledge staleness and reducing hallucinations. However, practical applications still face several challenges. First, there exists a knowledge conflict issue, where retrieved content may contradict internal model knowledge, leading to unfaithful or erroneous outputs \cite{zhang2025faithfulrag, xu2024parenting}. Second, noise interference and irrelevant information significantly degrade output quality due to redundant or misleading documents \cite{chang2024main, zhanglexical}. In multi-hop question-answering tasks, reasoning path deviation and cascading errors can lead to systemic failures in subsequent retrieval and generation steps \cite{li2024can, fang2025kirag}. Moreover, most RAG systems lack effective dynamic retrieval mechanisms, failing to determine when retrieval is necessary, which affects response efficiency \cite{guo2025dior, huanshuoctrla,liu2025hm}. Finally, knowledge boundary identification remains inadequate, as many models cannot effectively respond with ``I don’t know" when facing questions beyond their knowledge scope \cite{sun2025divide,huanshuoctrla}.

\subsection{Technological Pathways to Address RAG Challenges}
To tackle these challenges, recent studies have proposed various technical approaches. Preference modeling and gain assessment have emerged as promising strategies for improving paragraph filtering accuracy, exemplified by GainRAG, which quantifies the contribution of each paragraph to the correct answer \cite{jiang2025gainrag}. To address knowledge conflict and factual consistency issues, FaithfulRAG and RPO introduce self-knowledge verification mechanisms and reinforcement learning strategies, respectively, to enhance the reliability of generated results \cite{zhang2025faithfulrag,yan2025rpo}. In complex tasks like multi-hop QA, reasoning chain optimization and structured organization have become mainstream directions. RankCoT, DualRAG, and KiRAG improve reasoning capabilities through CoT ranking, dual-channel architecture, and iterative knowledge triplet utilization \cite{wu2025rankcot,cheng2025dualrag,fang2025kirag}. Meanwhile, document filtering and noise suppression mechanisms are widely explored, with MAIN-RAG introducing a multi-agent scoring filtering mechanism and Lexical Diversity-aware methods using contrastive learning to remove irrelevant content \cite{chang2024main,zhanglexical}. Furthermore, dynamic retrieval and trigger control mechanisms are evolving, with DioR proposing early and real-time classifiers to enable on-demand retrieval and significantly improve system responsiveness \cite{guo2025dior}.

\subsection{Reasoning Guidance and Knowledge Distillation}
An increasing number of studies are leveraging reasoning guidance and knowledge distillation to improve the intelligence of RAG systems. For example, Rationale Distillation utilizes LLMs to generate ``rationales" as signals for re-ranking documents and fine-tuning the reranker to better align with the generator's preferences \cite{jia2024bridging, meng2025ranking}. Memory-inspired iterative frameworks use multi-agent collaboration to integrate historical retrieval results, dynamically adjust queries, and filter noise, thus improving retrieval efficiency and quality \cite{qin2025towards}. Additionally, self-refinement mechanisms play a role in formalization tasks, with LTRAG building a thought-guided knowledge base to assist the formalization process and integrating symbolic solvers for dynamic optimization \cite{LTRAG}.

\subsection{Graph Enhancement and Structured Knowledge Modeling}
Structured knowledge sources, such as knowledge graphs and causal graphs, are playing an increasingly important role in RAG. SimGRAG transforms user queries into structured graph patterns and uses graph semantic distance to measure matching accuracy, thereby improving knowledge graph-driven retrieval effectiveness \cite{cai2024simgrag}. GEAR introduces a graph expansion mechanism (SyncGE) combined with a multi-step retrieval agent framework to support diverse path exploration in multi-hop QA tasks \cite{shen2024gear}. FRAG customizes retrieval workflows for queries of varying complexity, leveraging KG to provide explicit entity relationship support \cite{zhao2024frag}. CausalRAG further introduces causal graphs to guide information filtering during retrieval, enhancing contextual coherence and reasoning accuracy \cite{wang2025causalrag}. RAKG effectively addresses cross-document relation extraction and entity disambiguation by introducing a RAG-based document-level pre-entity retrieval and knowledge construction pipeline, enhancing knowledge graph accuracy \cite{zhang2025rakg}.These methods collectively push RAG toward structured knowledge fusion and graph-enhanced reasoning.

\subsection{Cross-modal and Fine-grained Retrieval}
As application scenarios expand, RAG is gradually extending into visual, tabular, and multi-modal domains. Fine-Grained VQA-RAG proposes a multi-modal knowledge unit (KU) combining images and textual descriptions to achieve zero-shot cross-modal retrieval and reasoning \cite{zhang2025fine}.VaLiK leverages vision-language models for image-text alignment and semantic verification, enabling unlabeled multimodal knowledge graph construction to enhance LLM reasoning \cite{liu2025aligning}. CoRE Framework focuses on structured data processing (e.g., tables), generating experience trajectories via Monte Carlo Tree Search (MCTS) and combining contrastive prompt design to enhance model reasoning \cite{gu2025toward}. Additionally, PIC chunking method proposes pseudo-instruction-based semantic segmentation without training, significantly optimizing document splitting and improving retrieval and QA performance \cite{DocumentSeg}. These studies mark RAG’s shift from single-text modalities toward multi-modal, fine-grained, and structured input.

\subsection{Interpretability, Honesty, and Controllable Generation}
To enhance transparency and reliability, multiple studies focus on interpretability, honesty, and controllable generation. Judge-as-a-Judge proposes filtering high-quality samples using judgment consistency for training, improving evaluation fairness \cite{liu2025judge}. CTRLA introduces representation-space intervention mechanisms, guiding honesty and monitoring confidence to make models more likely to acknowledge knowledge limitations and avoid false generation \cite{huanshuoctrla}. EXIT Mechanism proposes context-aware sentence-level compression to improve long-context processing efficiency while reducing interference \cite{hwang2024exit}. Additionally, Debate-Augmented RAG introduces a multi-agent debate mechanism to enhance objectivity and diversity in generated content \cite{hu2025removal}. These methods form the foundation for RAG’s evolution toward trustworthy AI, controllable generation, and ethical compliance .

\subsection{Application Expansion and Future Trends}
As RAG technology matures, its application scenarios are expanding into healthcare, education, law, and other fields. Medical and knowledge-intensive task applications have become hotspots, with RARE and TC–RAG achieving significant results in medical QA and commonsense reasoning \cite{wang2025rare,jiang2024tc}. Cross-model knowledge transfer and small-scale deployment have also become key directions, with DRAG and UniRAG promoting RAG in resource-constrained environments \cite{chen2025drag,UniRAG}. In terms of ethics and controllability, honesty-oriented controllable generation mechanisms are gaining attention, with Divide-Then-Align explicitly requiring models to be capable of responding with ``I don't know" to improve trustworthiness \cite{sun2025divide}. Additionally, human-machine collaborative and interactive retrieval-generation systems are emerging, with CR-Planner and ChainRAG supporting multi-turn dialogue and progressive reasoning to enhance user engagement \cite{li2024can,zhu2025mitigating}. Lastly, interpretability and traceability enhancements are becoming focal points, with SEAKR’s ``self-aware reranking" mechanism and KiRAG’s knowledge path recording helping improve model transparency and auditability \cite{yao2024seakr,fang2025kirag}.

\section{Methodolody}
 
    \begin{figure}[htp]
        \centering
        \includegraphics[width=0.98\linewidth]{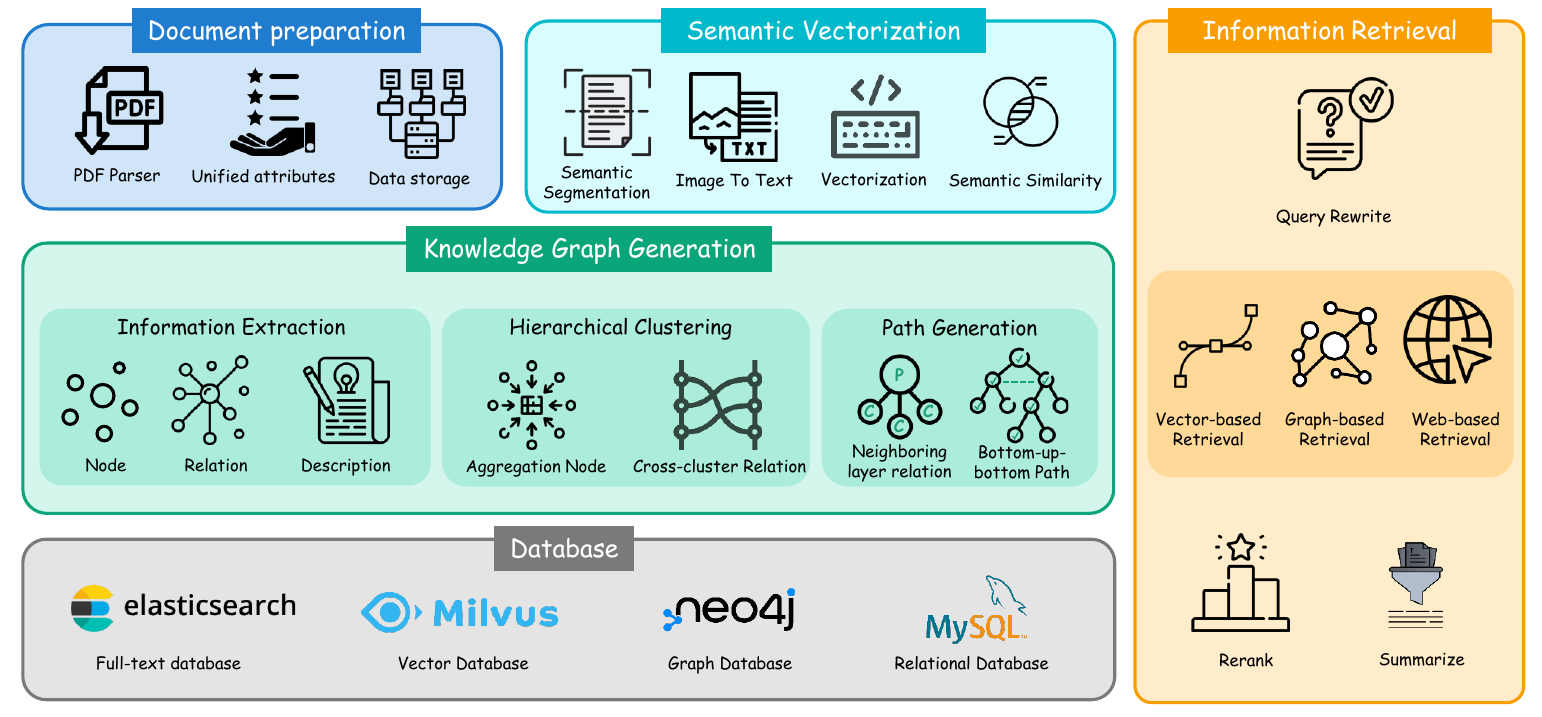}
        \caption{The Framework of the HetaRAG. Raw documents (PDF, image, web) are parsed into unified multimodal chunks. Each chunk is simultaneously indexed into four heterogeneous stores: Milvus (vector), Neo4j (graph), Elasticsearch (full-text), and MySQL (relational). At query time, the system rewrites the question, launches bottom-up graph traversal, semantic vector search, lexical full-text search, and SQL filtering in parallel, then reranks and summarizes the fused evidence to produce a grounded answer or a deep research report.}
        \label{fig:framework}
    \end{figure}

HetaRAG focuses on handling users’ complex question-answering tasks and generating in-depth research reports by leveraging heterogeneous knowledge stores.

\subsection{Data Processing}



In the HetaRAG system, data processing is a critical step that consists of two key phases: data parsing and data vectorization. 

In the data processing phase, unstructured files are transformed into semi-structured or structured data.  HetaRAG supports two batch data parsing methods: Docling \cite{Auer2024DoclingTR} and MinerU \cite{Wang2024MinerUAO}, offering flexible solutions to meet varying accuracy, structuring requirements, and document format compatibility.  MinerU is a layout-based PDF parsing method that focuses on extracting precise text blocks, tables, and images.  It analyzes the bounding box of each text block to determine the position and structure of content, preserving the layout, and providing detailed metadata for each element.  In contrast, Docling is tailored for parsing high-level document structures, making it particularly suitable for forms and reports with well-defined layouts.  These parsing methods facilitate the extraction of key elements from unstructured documents, converting them into structured data for further processing.


After data parsing, HetaRAG performs format conversion and content embedding on the parsed data.  For the different parsing results obtained from the two data parsing methods, HetaRAG has designed specific format conversion and embedding techniques to organize the data into unified fields. This standardization facilitates unified and precise matching and retrieval in the subsequent stages of the project.   Regarding embedding, HetaRAG supports two methods: text-only embedding using the BGE-m3 \cite{Chen2024BGEMM} model and multimodal embedding (text and image) using the QwenVL \cite{Zhang2024GMEIU} encoder.  Text embedding focuses on capturing the semantic representation of textual content, while multimodal embedding focuses on the unified encoding of both text and visual data.  Both embedding methods can be used by users in different application scenarios.  These operations lay the foundation for high-quality retrieval matching and efficient querying, ensuring optimal performance in later stages of the system.
	
\subsection{Database Construction}


To support the efficient storage and management of various data types, HetaRAG integrates four distinct database systems: Elasticsearch \cite{elasticsearch}, Milvus \cite{milvus}, MySQL \cite{mysql}, and Neo4j \cite{neo4j}. Each of these databases supports specific operations and is tailored to particular use cases, ensuring that HetaRAG can handle structured and unstructured data on a scale.


Elasticsearch is a distributed search engine designed for real-time retrieval and full-text search, supporting efficient document indexing and querying. It is used for keyword-based searches across large volumes of text data. Indexes documents in a structured format, enabling efficient real-time retrieval of relevant information. The project supports operations such as index creation, keyword search, and index deletion, as well as uploading and indexing of new data in PKL format. Once the index is established, users can perform fast and accurate search queries, retrieving optimal results based on specific keywords, thereby improving overall retrieval performance and user experience when handling large, unstructured datasets.


Milvus serves as a vector database, optimized for storing and performing similarity searches on high-dimensional embeddings. It is utilized to store embeddings generated during the data processing stage in a structured format, enabling vector similarity search across high-dimensional data.  The system supports operations such as collection creation, index addition, data insertion, and similarity querying, ensuring that the processed data is efficiently stored and retrieved.  By leveraging Milvus, HetaRAG facilitates rapid and accurate retrieval of semantically similar content, which is essential for handling large-scale, vectorized data and optimizing performance in tasks such as information retrieval and content matching.


MySQL is a relational database management system that facilitates the storage of structured data in tables, supporting complex queries such as filtering, aggregation, and joins. It is used to store structured data, particularly formatted data (e.g., tabular data), and supports operations such as database and table creation, data import, and complex query execution to enable efficient querying of structured data. By leveraging MySQL, HetaRAG ensures fast and reliable access to formatted data and facilitates complex analytical queries within the system.


Neo4j is a graph database that models data as nodes and relationships, offering powerful capabilities for representing and querying complex relational data.   It is used to store and manage graph-based data, enabling efficient querying of relationships between entities.   The system supports operations such as importing data, building graphs, and performing relationship-based searches.   By leveraging Neo4j, HetaRAG can efficiently uncover and analyze complex relationships within the data, making it an essential tool for tasks that require graph-based querying and relationship extraction.

\subsection{Knowledge-Graph Building}


HetaRAG supports two distinct knowledge graph construction methods: HiRAG \cite{huang2025retrieval} and LeanRAG \cite{zhang2025leanrag}, which are designed to enhance the integration of structured knowledge into the RAG pipeline and improve the system’s reasoning capabilities over complex data.  HiRAG emphasizes hierarchical knowledge aggregation by constructing multi-level entity representations, thereby enabling more effective reasoning and retrieval across different levels of semantic granularity.  LeanRAG focuses on generating rich entity-relation triples and organizing them into a comprehensive knowledge graph via a dynamic, multi-level hierarchy.  Both approaches contribute to building coherent and scalable knowledge representations, ensuring that HetaRAG can process domain-specific information with higher efficiency and precision.


For both HiRAG and LeanRAG construction, HetaRAG abstracts the process into three stages: (1) extraction of entity-relation triples, (2) knowledge graph construction, and (3) graph-based retrieval and question answering.   In the first stage, two triple extraction methods are provided: one based on CommonKG and the other on GraphRAG.   The CommonKG approach leverages large-scale commonsense knowledge patterns to extract semantically meaningful triples from unstructured text.   In contrast, the GraphRAG method employs LLM-guided extraction tailored for retrieval-augmented generation, enabling more context-aware representations.   In the second stage, each method constructs knowledge graphs according to its respective paradigm.   HiRAG integrates knowledge hierarchically through an unsupervised layered indexing mechanism and novel bridging strategy, while LeanRAG utilizes a hierarchical graph aggregation approach to abstract fine-grained entities and relations into a semantically rich, navigable network.   These graphs are subsequently used for downstream graph-based retrieval and reasoning tasks.

\subsection{Advanced Retrieval}


\textbf{Hybrid Retrieval} in HetaRAG combines vector-based retrieval(Milvus), with keyword-based search(Elasticsearch).  This approach enables more accurate and comprehensive search results by integrating the strengths of both retrieval methods.  The system utilizes a parameter, alpha, to control the weight of each retrieval method, allowing for fine-tuning of the balance between vector search and keyword search.  By merging the semantic power of vector-based search and the precision of keyword matching, Hybrid Retrieval ensures that relevant content is effectively retrieved, optimizing performance across a wide range of queries and enhancing retrieval accuracy.


\textbf{Reranking} in HetaRAG provides a flexible reranking framework that further optimizes search results through reranking of retrieved content.  The system supports various reranking techniques, such as vLLM deployment, Hugging Face model downloads, and direct API calls.  The reranking process allows for the selection of the most relevant results based on additional context and relevance, thereby improving the quality of the retrieved content.  This flexibility enables users to choose from a variety of reranking models, whether downloaded from external repositories or deployed locally.  By supporting parent document retrieval and vector-based database systems like Milvus, the reranking mechanism enhances the accuracy and relevance of the final search results, fostering more precise and customized outcomes.

\subsection{DeepSearch}

    \begin{figure}[htp]
        \centering
        \includegraphics[width=0.98\linewidth]{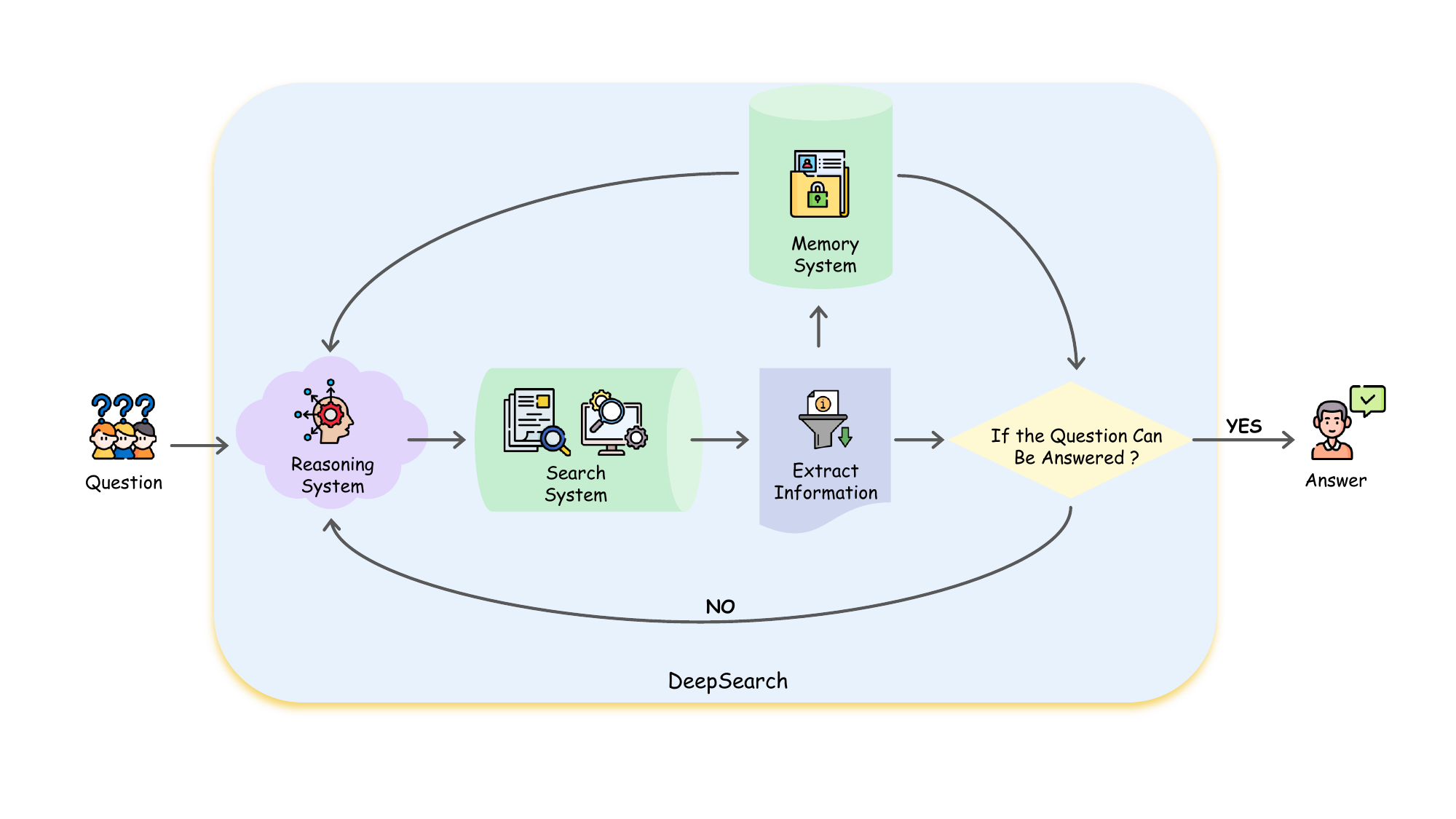}
        \caption{The Framework of DeepSearch}
        \label{fig:DeepSearch-framework}
    \end{figure}


The DeepSearch component in HetaRAG is designed to address complex queries that require multi-step retrieval and iterative reasoning across diverse information sources. This component integrates semantic retrieval, information extraction, and critical reasoning to iteratively gather and synthesize information, ultimately providing accurate answers.  The process is structured in multiple reasoning steps, where each stage refines the search and reasoning process based on intermediate results. 


The core of the system is the MultiHopAgent—a structured, iterative retrieval-and-reasoning agent whose design is adapted from the WebWalker framework proposed by  \cite{wu2025webwalker}.
The overall framework is illustrated in Figure \ref{fig:DeepSearch-framework}. Initially, the agent rewrites the query to refine its scope and context.  Based on the rewritten query, the system performs a search to retrieve relevant documents.  The retrieved content is then analyzed, and key information is extracted.  The system compares the extracted information with the original query to assess whether the current content is sufficient to answer the question.  If the information is inadequate, the system generates new queries derived from the search results and restarts the retrieval process.  This cycle of search, analysis, comparison, and refinement continues until the system generates a comprehensive and accurate answer to the original question.  Through this multi-step approach, the MultiHopAgent ensures that the generated answer is well-supported by relevant and contextually consistent information.


The retrieval process in MultiHopAgent supports multiple data sources, such as pre-constructed knowledge bases and web pages, ensuring that the system can retrieve contextual information from diverse repositories.    By integrating various search mechanisms, the multi-hop reasoning framework provides a robust solution for addressing complex queries that require the synthesis of information from multiple domains.

\subsection{DeepWriter}

\begin{figure}[htp]
        \centering
        \includegraphics[width=0.98\linewidth]{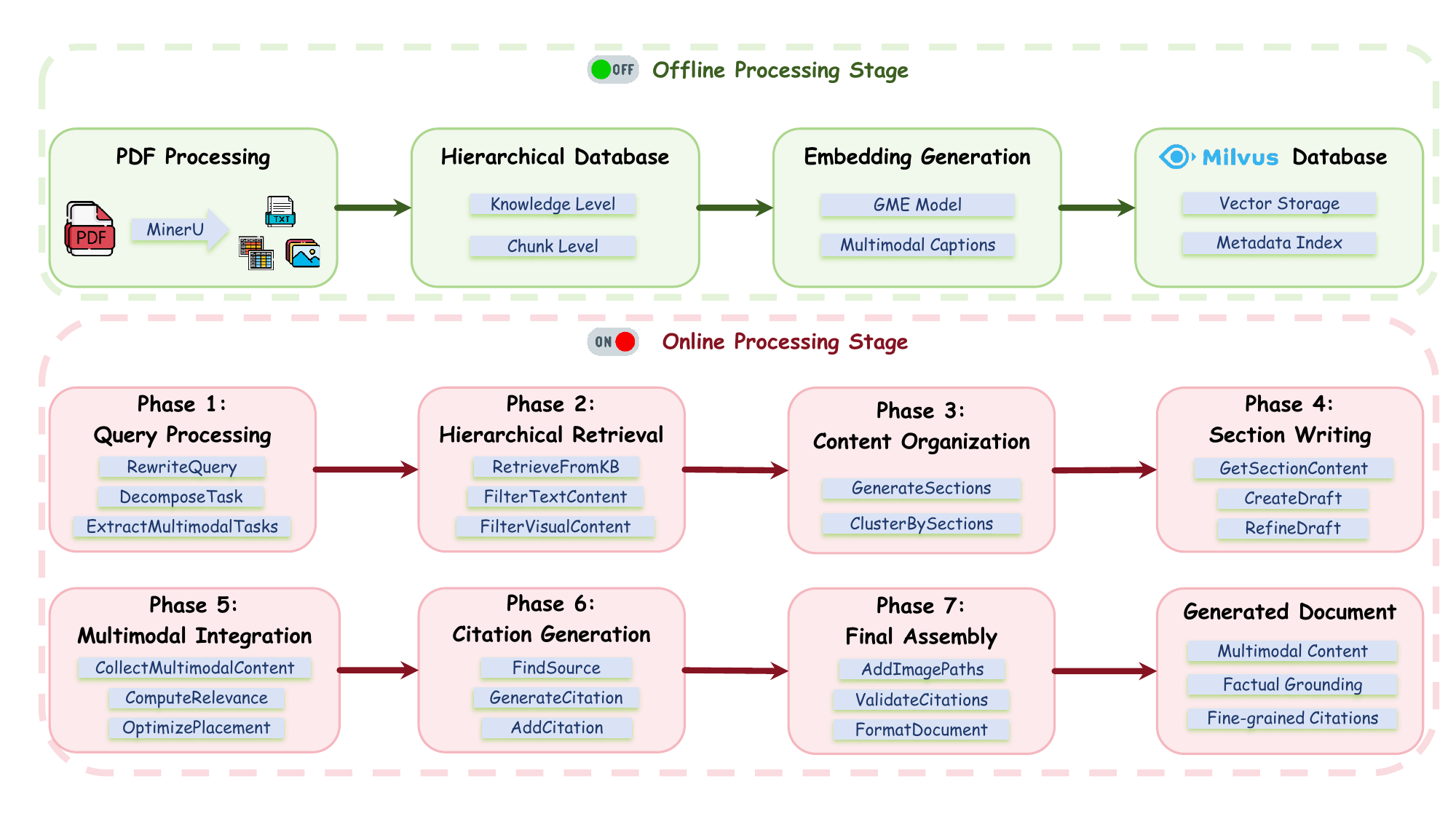}
        \caption{The Framework of DeepWriter}
        \label{fig:DeepWriter-framework}
    \end{figure}

The Multimodal Report Generation component in the HetaRAG system enables the automatic generation of reports from unstructured documents, and it is implemented by DeepWriter \cite{mao2025deepwriterfactgroundedmultimodalwriting}.  DeepWriter leverages large language models (LLMs) and a vector database to extract semantic information from unstructured documents and generate structured textual reports conditioned on user queries. The overall pipeline is illustrated in Figure \ref{fig:DeepWriter-framework}.  The system allows users to input queries, which then drive the report generation process, ensuring that the resulting report is comprehensive, coherent, and relevant to the user's specific needs. This capability is particularly beneficial for generating professional-grade, factually grounded documents, as it integrates both textual and visual data. 

The report generation process is divided into several stages.     First, DeepWriter retrieves contextual information through DocFinder by leveraging vector search, ensuring that the retrieved content is relevant and semantically aligned with the query.     It then integrates this content, combining the query-driven information with the LLM to generate the report.    Visual elements like charts, tables, and images are placed within the report based on their relevance to the text, optimizing the integration of multimodal data.     Finally, the ReportProcessor ensures that the generated document is formatted in Markdown and includes all necessary citations.     The report is thoroughly fact-checked, with each claim or visual element accurately cited, providing transparency and verifiability.     This robust framework enhances the quality and reliability of the generated reports, supporting complex tasks across diverse domains.

\section{Experiment}

The experimental evaluation of the HetaRAG system is divided into three main parts, each targeting a different aspect of the system's capabilities. Section 3.1, RAG-Challenge Evaluation, focuses on assessing the retrieval results. Section 3.2 evaluates the effectiveness of multi-hop reasoning when handling complex questions that require reasoning across multiple information sources. Section 3.3 is dedicated to evaluating the quality of multimodal report generation, examining how well the system generates structured reports that integrate both textual and visual content. Each section provides valuable insights into the strengths and potential improvements of the system.

\subsection{RAG-Challenge}

\subsubsection{Experimental Setup}

\textbf{Dataset.} To assess the retrieval and reasoning capabilities of HetaRAG, we adopt the dataset and evaluation protocol from the Enterprise RAG Challenge Round 2\cite{RAGChallengedata}. The benchmark comprises domain-specific queries paired with gold-standard answers annotated by experts, along with relevant document contexts.Evaluation follows the RAG Challenge metrics, including the Retrieval Score (R) and Generation Score (G) and Score = R/3 + G (max 133), reflecting both retrieval accuracy and answer generation quality.


\textbf{Implementation Details.}  We conduct experiments using the pipeline designed by Ilya Rice for this dataset, which incorporates Docling for PDF parsing, BGE-m3 for data embedding, and Milvus for constructing the vector database. Multiple large language models, such as Qwen2.5-72B and GPT-4o, are employed for answer generation. Additionally, the system supports various large model frameworks, including API keys, vLLM, and Ollama.

\subsubsection{Result}

Table~\ref{tab:RAG-Challenge Result} presents the experimental results of the HetaRAG system on the RAG-Challenge dataset, highlighting the effects of reranking strategies and query rewriting across two large language models: ChatGPT-4o and Qwen2.5-72B-Instruct. Among all configurations, ChatGPT-4o combined with the bge-reranker-large achieved the highest overall score (117.0), demonstrating that reranking significantly enhances both retrieval (R) and generation (G) performance. 

For Qwen2.5-72B-Instruct, the best result (109.8) was also obtained using reranking with bge-reranker-v2-gemma, outperforming its baseline configuration by over 3 points. These findings confirm that rerankers contribute more substantially to performance gains than query rewriting and that the choice of reranker model affects overall effectiveness. The results also suggest that, while compact models like Qwen2.5 perform competitively, ChatGPT-4o maintains a notable advantage in generation quality when enhanced with proper reranking strategies.

\begin{table}[htbp]
\centering
\caption{RAG-Challenge Result}
\label{tab:RAG-Challenge Result}
\begin{tabular}{lllccc}
\toprule
\textbf{Model} & \textbf{Rerank} & \textbf{QueryRewrite} & \textbf{R} & \textbf{G} & \textbf{Score} \\
\midrule
{ChatGPT-4o} 
& - & - & 78.3 & 73.8 & 113.0 \\
& - & $\surd$ & 78.9 & 72.8 & 112.3 \\
& bge-reranker-large & - & 79.7 & 77.2 & 117.0 \\
& bge-reranker-v2-gemma & - & 79.7 & 74.8 & 114.7 \\
& bge-reranker-v2-m3 & - & 78.7 & 73.8 & 113.2 \\
\midrule
\addlinespace
{Qwen2.5-72B-Instruct}
& - & - & 77.1 & 70.0 & 108.6 \\
& - & $\surd$ & 76.8 & 67.8 & 106.2 \\
& bge-reranker-large & - & 77.5 & 68.6 & 107.3 \\
& bge-reranker-v2-gemma & - & 78.2 & 70.7 & 109.8 \\
& bge-reranker-v2-m3 & - & 77.8 & 69.6 & 108.5 \\
\bottomrule
\end{tabular}
\end{table}

\subsection{DeepSearch}





\textbf{Dataset.}  To support qualitative evaluation of the DeepSearch module in the HetaRAG framework, we construct a synthetic DeepSearch QA dataset   that emphasizes deep semantic search and compositional reasoning.  Each query is designed to require multi-step retrieval, cross-document inference, and entity disambiguation, closely reflecting real-world knowledge-intensive tasks.  The accompanying document set spans diverse but interconnected facts, enabling controlled testing of iterative querying, memory accumulation, and query rewriting.

\begin{figure}[htp]
        \centering
        \includegraphics[width=0.98\linewidth]{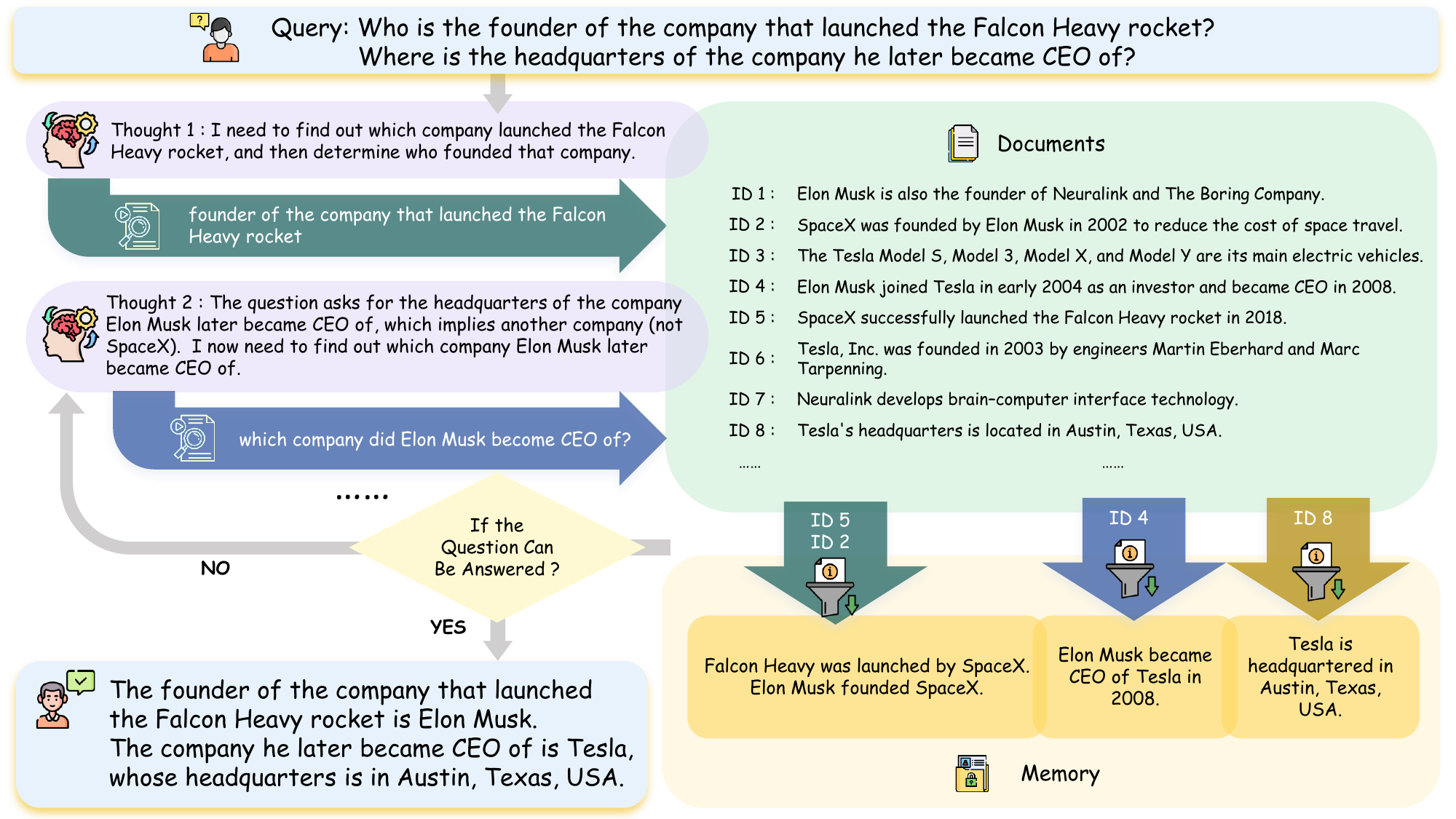}
        \caption{The Example of DeepSearch}
        \label{fig:example_DeepSearch}
    \end{figure}

\textbf{Result.} As shown in Figure \ref{fig:example_DeepSearch}, the DeepSearch module conducts deep iterative search by decomposing the query into focused sub-questions.  It effectively leverages retrieved information to update memory, enabling cross-document reasoning and evidence accumulation.  This result demonstrates DeepSearch’s ability to perform deep semantic search and dynamically utilize intermediate findings to construct accurate, well-supported answers.

\subsection{Deepwriter}

\subsubsection{Experimental Setup}
\textbf{Dataset.} To evaluate the report generation capability of HetaRAG’s DeepWriter module, we use the World Trade Report (WTR) dataset, which consists of 23 annual reports published by the World Trade Organization from 2001 to 2024. These documents serve as the offline corpus for generating long-form articles in response to user queries. For evaluation, following \cite{shao2024assisting,jiang2024into,xi2025omnithink}, we adopt Prometheus2-7B\cite{kim2024prometheus} as the evaluation model to assess the quality of generated articles. The evaluation follows four dimensions—Interest Level, Coherence and Organization, Relevance and Focus, and Broad Coverage—each scored on a 1–5 scale.  These criteria comprehensively measure the effectiveness of DeepWriter in producing informative, coherent, and engaging reports grounded in domain-specific knowledge.

\textbf{Implementation Details.} We employ MinerU for PDF parsing, storing the preprocessed data in Milvus along with embeddings generated by GME (gme-Qwen2-VL-2B-Instruct). Visual elements such as images and tables are captioned using Qwen2.5-VL-7B, an advanced vision-language model. Qwen2-7B serves as the core language model for executing the report generation tasks.

\subsubsection{Result}

The performance of DeepWriter is evaluated on the WTR dataset and compared against three baselines: (1) Long-context LLMs such as Qwen-Plus with a 131K token context window; (2) Naive RAG, which retrieves documents based on the original query without structured planning; and (3) specialized long-form writing systems including STORM and CO-STORM. As shown in Table~\ref{tab:system_config}, DeepWriter demonstrates competitive results, particularly in coherence and organization, highlighting its ability to generate structured long-form content even with compact models.

\begin{table}[htbp]
\centering
\caption{Configuration for different systems}
\label{tab:system_config}
\begin{tabular}{lccccc}
\toprule
 & \textbf{Qwen-Plus} & \textbf{Qwen-Plus (w titles)} & \textbf{STORM} & \textbf{CO-STORM} & \textbf{DeepWriter} \\
\midrule
\textbf{Model} & Qwen-Plus & Qwen-Plus & GPT-4o & GPT-4o & Qwen2-7B \\
\textbf{Source} & Internal & Internal & search & search & offline KB \\
\textbf{Generation} & single turn & single turn & multi-turn & multi-turn & multi-turn \\
\textbf{Web search} & no & no & yes & yes & no \\
\bottomrule
\end{tabular}
\end{table}

\begin{table}[htbp]
\centering
\caption{Average Performance Scores Across All Dimensions}
\label{tab:deepwriter_result}
\begin{tabular}{p{2cm}ccccc} 
\toprule
 & \textbf{Qwen-Plus} & \textbf{Qwen-Plus (w titles)} & \textbf{STORM} & \textbf{CO-STORM} & \textbf{DeepWriter} \\
\midrule
\textbf{Ave Score} & 4.92 & 4.88 & 4.88 & 4.68 & 4.64 \\
\bottomrule
\end{tabular}
\end{table}
    
As illustrated in Table~\ref{tab:deepwriter_result}, DeepWriter achieves performance comparable to GPT-4o-based systems, validating the effectiveness of combining semantic retrieval and structured generation. Notably, incorporating title planning from DeepWriter into Qwen-Plus improves coherence, emphasizing the benefits of decomposition and guided generation. Nonetheless, DeepWriter underperforms in certain dimensions, such as Coverage and Relevance, underscoring the limitations of smaller models compared to larger LLMs in complex content generation scenarios.


\section{Conclusion}
We introduce HetaRAG, a deep hybrid retrieval-augmented generation (RAG) framework that operates over heterogeneous data stores to answer complex questions from existing knowledge bases and produce in-depth, illustrated research reports. This technical report summarizes our preliminary exploratory work and releases the corresponding open-source code. In future work, we will construct a multimodal hybrid-retrieval solution anchored in a knowledge graph that seamlessly unites relational, vector and full-text search databases.


\clearpage


\bibliography{example}  

\end{document}